\documentclass[apj]{emulateapj}
\usepackage{amsmath}                
\usepackage{amsfonts}               
\usepackage{amssymb}                
\usepackage{epsfig}                 
\usepackage[para,online,flushleft]{threeparttable}

\newcommand{\cm}{{~\rm cm}}
\newcommand{\km}{{~\rm km}}
\newcommand{\s}{{~\rm s}}

\newcommand{\G}{{~\rm G}}

\newcommand{\erg}{{~\rm erg}}

\newcommand{\zams}{\mathrm{ZAMS}}
\newcommand{\nar}{{~\rm New Astronomy Reviews}}
\newcommand{\na}{{~\rm New Astronomy}}

\begin{document}

\title{Storing magnetic fields in pre-collapse cores of massive stars}

\author{Inbal Peres\altaffilmark{1}, Efrat Sabach\altaffilmark{1} \& Noam Soker\altaffilmark{1,2}}

\altaffiltext{1}{Department of Physics, Technion -- Israel Institute of Technology, Haifa
32000, Israel}

\altaffiltext{2}{Guangdong Technion Israel Institute of Technology, Shantou, Guangdong Province 515069, China}

\thanks{E-mail: soker@physics.technion.ac.il, efrats@physics.technion.ac.il}

\begin{abstract}
We argue that the radiative zone above the iron core in pre-collapse cores of massive stars can store strong magnetic fields. To reach this conclusion we use the stellar evolutionary code MESA to simulate the evolution of two stellar models with initial masses of $M_{\rm ZAMS} =15 M_\odot$ and $M_{\rm ZAMS} =25 M_\odot$, and reveal the entropy profile above the iron core just before core collapse. Just above the iron core there is a thin zone with convective shells. We assume that a dynamo in these convective shells amplifies magnetic fields and forms magnetic flux loops.
By considering the buoyancy of magnetic flux loops we show that the steep entropy rise in the radiative zone above the dynamo can prevent buoyancy of flux loops with magnetic fields below about ${\rm several} \times 10^{12} \G$. 
When this radiative zone collapses on to the newly born neutron star the converging inflow further amplifies the magnetic fields by a factor of about a hundred. After passing through the stalled shock at about a hundred kilometres from the center, these strong magnetic fields together with instabilities can facilitate the launching of jets that explode the star in the frame of the jittering jets explosion mechanism.
Our study further supports the claim for the necessity to include magnetic fields in simulating the explosion of CCSNe.
\end{abstract}


\section{Introduction}
\label{sec:intro}

Analysis of observations of core collapse supernovae (CCSNe) and CCSN remnants suggests that jets play a major role in the explosion of many, or even all, CCSNe (e.g., \citealt{Gonzalezetal2014, Marguttietal2014, Milisavljevicetal2015, FesenMilisavljevic2016, Inserraetal2016, Bearetal2017, Mauerhanetal2017, GrichenerSoker2017, BearSoker2017, Garciaetal2017, Marguttietal2017a, Piranetal2017, Tanakaetal2017, LopezFesen2018}, limiting to some papers from the past five years). 
Most theoretical studies that involve jets in the last two decades (e.g. \citealt{Khokhlovetal1999, MacFadyen2001, Hoflich2001, Woosley2005, Burrows2007, Couch2009, Couch2011, TakiwakiKotake2011, Lazzati2012, Maedaetal2012, Mostaetal2014, Nishimura2015, BrombergTchekhovskoy2016, Gilkis2018, Nishimuraetal2017, Sobacchietal2017}) assume that some special conditions are required for jets to play a role in the explosion, in particular that the pre-collapse core must be rapidly rotating. This condition is rarely met though, as it requires a stellar binary companion to directly spin-up the core of the CCSN progenitor. 

In contrast, the jittering jets explosion mechanism posits that jets explode all CCSNe (e.g., \citealt{PapishSoker2011, Papishetal2015, GilkisSoker2015, Gilkisetal2016}; review by \citealt{Soker2016Rev}). In the case of rapidly rotating pre-collapse cores the accretion flow forms an accretion disk around the newly born neutron star, and this Keplerian accretion disk launches the jets while maintaining a constant jets' direction. The challenge is to launch jets when the specific angular momentum of the accretion gas is too low to form a Keplerian accretion disk (i.e., one that is supported against gravity by the centrifugal force alone). In those cases accretion of stochastic angular momentum leads to the launching of jittering jets. The seeds of the stochastic angular momentum come from the random convective motion in the core or the envelope of the star (e.g., \citealt{GilkisSoker2014, GilkisSoker2015, GilkisSoker2016, Quataertetal2019}). 

Although the neutrino explosion mechanism can form non-spherical structures (e.g., \citealt{Wongwathanaratetal2015}), in the case of the asymmetrical structures of SN~1987A (e.g., \citealt{Larssonetal2019}) these models have problems to explain all morphological features (e.g., \citealt{Soker2017RAA, Abellanetal2017}). \cite{BearSoker2018} argue that the jittering jets explosion mechanism can better fit the morphology of SN~1987A. 

In the most recent version of the jittering jets explosion mechanism \citep{Soker2019}, in the case of low angular momentum of the accreted gas jets are launched by the spiral standing accretion shock instability. This instability leads to a stochastically variable angular momentum of the accreted gas that results in alternating shear zones in the inflow around the newly born neutron star (e.g., \citealt{BlondinShaw2007, Rantsiouetal2011, Fernandez2010, Endeveetal2012, Hankeetal2013, GuiletFernandez2014, Iwakamietal2014, Fernandez2015,  Blondinetal2017, Kazeronietal2017, OConnorCouch2018}). The shear amplifies the magnetic fields that are thought to be crucial for the launching of jets. The jets are intermittent and have variable directions, i.e., jittering. Neutrino heating is an important ingredient in further energizing the jets. According to the jittering-jets explosion mechanism these jets are the first to break through the stalled accretion shock, and then explode the star. In other words, the jittering jets locally revive the stalled accretion shock along the momentarily polar directions, and by that they explode the star.

Strong magnetic fields in the pre-collapse core will result in strong magnetic fields in the spiral standing accretion shock instability zone at explosion, both because of the spiral standing accretion shock instability and turbulence (e.g., \citealt{Mostaetal2015}). 
\cite{Wheeleretal2015} mention the amplification of magnetic fields by the shear in the pre-collapse core. They suggest that magnetic fields of about $10^{12} \G$ might exist at the edge of the pre-collapse iron core.
\cite{Zilbermanetal2018} evolve single stellar models and find large rotational shear above the pre-collapse iron core. This shear, they argue, amplifies magnetic fields that are further amplified after collapse, and play a crucial role in launching jets within the jittering jets explosion mechanism. 
The conclusion from the above studies and others (e.g., \citealt{Hegeretal2005}) is that the shear in the pre-collapse core amplifies magnetic fields. 

In the present study we examine the possibility to store magnetic energy in the radiative zones of the pre-collapse inner core. In section \ref{sec:storing} we describe our assumptions to store magnetic energy. In section \ref{sec:setup} we describe the numerical procedure for the stellar evolution code, and in section \ref{sec:results} we describe the results.  In section  \ref{sec:assumptions} we examine some of our simplifying assumptions. Our short summary is in section \ref{sec:summary}. 

\section{Storing magnetic fields in radiative zones}
\label{sec:storing}

There are different studies of the buoyant rise of magnetic flux tubes through the radiative zones of stars (e.g., \citealt{MacGregorCassinelli2003, MacDonaldMullan2004}). We will use the terms loop and tube to have the same meaning. Mathematically we treat a tube, but to emphasise the nature of the magnetic field lines we should keep in mind that the tube closes on itself to form a loop. 
We follow the analysis of \cite{HarpazSoker2009} which is most relevant to our goals, and
present here the essential derivation of the magnetic field that can be stored in a radiative zone.  
 
\begin{figure}  %
\vskip -2.0 cm
\hskip -2.0 cm
\begin{tabular}{cc}
{\includegraphics[scale=0.56]{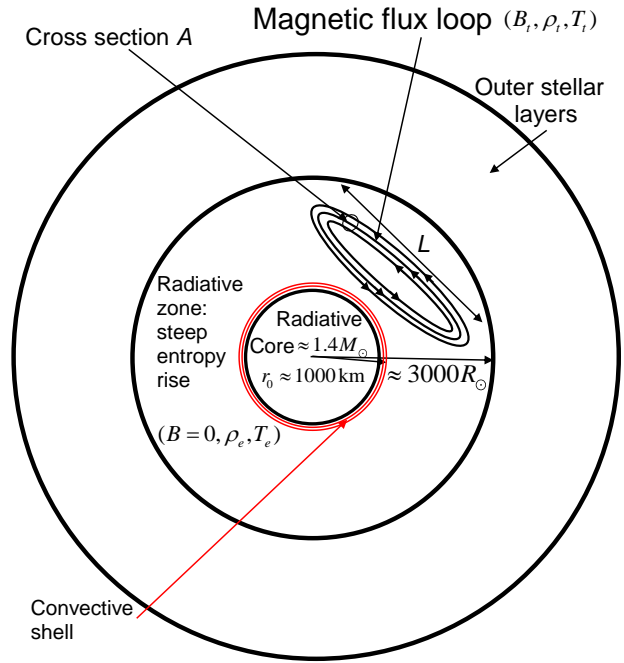}}
\end{tabular}
\vskip -3.9 cm
\caption{A schematic drawing (not to scale) of our $M_{\rm ZAMS} = 15 M_\odot$ stellar model just before core collapse. A schematic  flux loop is embedded in a radiative zone with an average steep entropy rise. }
\label{fig:flux}
\end{figure}

We study a magnetic flux tube that is formed by a dynamo at the base of the radiative zone. 
The convection zone below the radiative zone and the large shear between the convective and radiative zones (e.g., \citealt{Zilbermanetal2018}) can form an efficient dynamo. 
Let the bottom of the radiative zone be at a radius $r=r_0$, and let the initial magnetic field intensity and density inside the flux loop be $B_{t0}$ and $\rho_{t0}$, respectively. The cross-section of the tube is $A(r)$, with $A_0 \equiv A(r_0)$. We will consider a segment of the loop, a tube, of length $L$. We also assume that at $r=r_0$ the temperature of the loop equals to that of the environment temperature $T_{t0}=T_e(r_0)\equiv T_{e0}$. 
The environment density is $\rho_e(r)$, and we assume that the magnetic field in the environment is much weaker than that inside the loop. The environment pressure is $P_e(r)=\rho_e k T_e/\mu m_H$, where $\mu m_H$ is the mean mass per particle.
We present a schematic drawing of the flux loop in Fig. \ref{fig:flux}.

We define the initial ratio of thermal to magnetic pressure inside the loop
\begin{equation}
\beta_0 \equiv \left( \frac {P_{\rm thermal}}{P_B} \right)_{t0}=\frac{\rho_{t0} k T_{t0}}{\mu m_H}
\left(  \frac{B_{t0}^2}{8 \pi}  \right)^{-1}.
\label{eq:beta}
\end{equation}
The total pressure inside the loop reads
\begin{equation}
P_t={P_{\rm thermal}}+{P_B}=\frac{\rho_t k T_t}{\mu m_H}+ \frac{B_t^2}{8 \pi}.
\label{eq:ptube1}
\end{equation}
In what follows we neglect the change in the molecular weight $\mu$, although it might change a little inside the core. We discuss this point in section \ref{sec:assumptions}. 

We take the magnetic field inside the loop to evolve according to flux conservation
\begin{equation}
B_t=B_{t0} \frac{A_0}{A}=B_{t0} \left( \frac{\rho_t}{\rho_{t0}} \right)^{\delta},
\label{eq:btube}
\end{equation}
where the value of $\delta$ depends on the relative variation of $A$ and $L$.
If the magnetic field is strong then magnetic tension preserves the length $L$, and  $\delta \simeq 1$. 
For a weak field that is dragged by the flow we expect a more or less random field for which $\delta \simeq 2/3$. We take here $\delta=2/3$ as we start with weak magnetic fields.
We assume that the loop evolves adiabatically such that its temperature is $T_t \propto \rho_t^{\gamma_{\rm ad}-1}$. With the adiabatic assumption and with equation (\ref{eq:btube}), equation (\ref{eq:ptube1}) reads 
\begin{equation}
P_t=\frac{\rho_{t0} k T_{t0}}{\mu m_H} \left( \frac{\rho_t}{\rho_{t0}} \right)^{\gamma_{\rm ad}} +
\frac{B_{t0}^2}{8 \pi} \left( \frac{\rho_t}{\rho_{t0}} \right)^{2 \delta}.
\label{eq:ptube2}
\end{equation}
The adiabatic index varies with radius, but in the relevant zones we can take it to be  
$\gamma_{\rm ad}\simeq 4/3$. 

By using the definition of $\beta_0$ we rewrite equation (\ref{eq:ptube2}) in the form 
\begin{equation}
P_t=\frac{P_{e0}}{1+\beta_0^{-1}} \left[ \left( \frac{\rho_t}{\rho_{t0}} \right)^{\gamma_{\rm ad}} +
\frac{1}{\beta_0} \left( \frac{\rho_t}{\rho_{t0}} \right)^{2 \delta} \right],
\label{eq:ptube3}
\end{equation}
where from pressure equilibrium between the tube (loop) and the environment we used $P_{e0}=P_{t0}=(1+\beta^{-1}_0)\rho_{t0} k T_{t0}/\mu m_H$.
Using our assumption that the temperature of the tube and of the environment are equal at the origin of the tube, we find the following expression for the density of the tube at its origin $\rho_{t0}=\rho_{e0}(1+\beta^{-1}_0)^{-1}$.
The total pressure in the tube equals that of the environment as it rises through the radiative zone $P_t=P_e(r)$.
We use these expressions for $\rho_{t0}$ and $P_t$ to cast equation (\ref{eq:ptube3}) into 
\begin{equation}
\begin{split}
\frac{P_e(r)}{P_{e0}} = &
 \left( \frac{\rho_t}{\rho_{e0}} \right)^{\gamma_{\rm ad}}(1+\beta_0^{-1})^{\gamma_{\rm ad}-1} \\ &
+
\frac{1}{\beta_0} \left( \frac{\rho_t}{\rho_{e0}} \right)^{2 \delta}(1+\beta_0^{-1})^{2 \delta-1}  .
\end{split}
\label{eq:ptube4}
\end{equation}

As the density inside the flux tube is lower than that in the environment it buoyantly rises. Let us examine the case where the tube reaches an equilibrium position where it does not rise anymore. Namely, its density becomes equal to the environment density $\rho_t=\rho_e(r)$.
Substituting this equality in equation (\ref{eq:ptube4}), and multiplying by
$[\rho_{e0}/\rho_e(r)]^{\gamma_{\rm ad}}$, we find
\begin{equation}
\begin{split}
& \frac{ P_e(r)/[\rho_e(r)]^{\gamma_{\rm ad}}}{ P_{e0}/\rho_{e0}^{\gamma_{\rm ad}}  }
 =\left( {1+\beta_0^{-1}} \right)^{\gamma_{\rm ad}-1}
\\ &
+
\left( {1+\beta_0^{-1}} \right)^{2\delta-1} \frac{1}{\beta_0}
 \left[ \frac{\rho_e(r)}{\rho_{e0}} \right]^{2 \delta -\gamma_{\rm ad}}.
\end{split}
\label{eq:equilibrium1}
\end{equation}

To reveal the nature of equation (\ref{eq:equilibrium1}) we take the entropy in the environment as $S_e(r)\simeq S_\gamma(r) \equiv P_e / \rho_e^{\gamma_{\rm ad}}$, so the left hand side of that equation is $\simeq S_e(r)/{S_{e0}}$. We can further simplify that equation if we take the approximation $\delta=2/3$ and $\gamma_{\rm ad} \simeq4/3$. For the $M_{\rm ZAMS}=15 M_\odot$  and $M_{\rm ZAMS}=25 M_\odot$ models that we simulate we find that in the relevant mass zone $1 M_\odot < m < 2 M_\odot$ the adiabatic index is $1.34 \la \gamma_{\rm ad} \la 1.44$ and $1.35 \la \gamma_{\rm ad} \la 1.36$, respectively.  
Equation (\ref{eq:equilibrium1}) for the radius at which the magnetic flux tube/loop comes to rest reads 
\begin{equation}
\frac{S_e(r)}{S_{e0}} \simeq 
\frac{S_\gamma(r)}{S_{\gamma 0}} \simeq
\left( {1+\beta_0^{-1}} \right)^{1.3}. 
\label{eq:equilibrium2}
\end{equation}
Equation (\ref{eq:equilibrium2}) implies that buoyantly rising magnetic flux tubes will stop their rise, namely be `stored', in the radiative zone if the initial magnetic field at dynamo location, $B_{t0}$ (given by equation \ref{eq:beta}), is smaller than an upper value of $B_{\rm max, 0}$. The upper value is derived from equation (\ref{eq:equilibrium2}) and reads  
\begin{equation}
\begin{split}
& \left( {1+\beta_{\rm max,0} ^{-1}} \right)^{1.3} = 
\left[ 1+ \left( \frac{\rho_{t0} k T_{t0}}{\mu m_H}\right)^{-1}
  \frac{B_{\rm max, 0}^2}{8 \pi}   \right]^{1.3}
\\ &
\simeq \frac{S_\gamma(r_{\rm max})}{S_{\gamma 0}} \simeq 
\frac{S_e(r_{\rm max})}{S_{e0}} . 
\end{split}
\label{eq:equilibrium3}
\end{equation}
If the magnetic field in the flux loop does not obey the relation $B_{t0} \la B_{\rm max, 0}$ the flux loop will exit the outer boundary of the radiative zone and continue to rise, and will not be accreted. 

The full stability condition, namely that the flux loop is in equilibrium, is more complicated.  
\cite{MacGregorCassinelli2003} show that radiative diffusion and rotation might make flux loops that are stable to radial perturbations to become unstable to perturbations in other directions. However, in that case the motion of the flux loop out from the radiative zone is long as it proceeds on a radiative diffusion timescale.  
For the goals of the present study we can stay with the approximate condition (\ref{eq:equilibrium3}).  

\section{Stellar properties}
\label{sec:setup}

We construct a set of stellar models using Modules for Experiments in Stellar Astrophysics (MESA, version 10398; \citealt{Paxton2011,Paxton2013,Paxton2015}), with initial masses of $M_\zams=15 M_\odot$ and $M_\zams=25 M_\odot$, with an initial metallicity of $Z=0.019$. In this study we do not consider rotating cores since our aim is to study the implications to the jittering jets explosion mechanism, and this mechanism was constructed to work in particular when the core does not rotate or rotates very slowly. 

Only the very last evolutionary phase before core collapse is relevant to our study, and so we present the stellar structure when the iron core is massive.
We are also interested in the magnetic fields in the mass that will be the last to be accreted by the newly born NS. For that we zoom in to the core zone. 

\begin{figure}[ht]
\hskip -0.20 cm
{\includegraphics[scale=0.46]{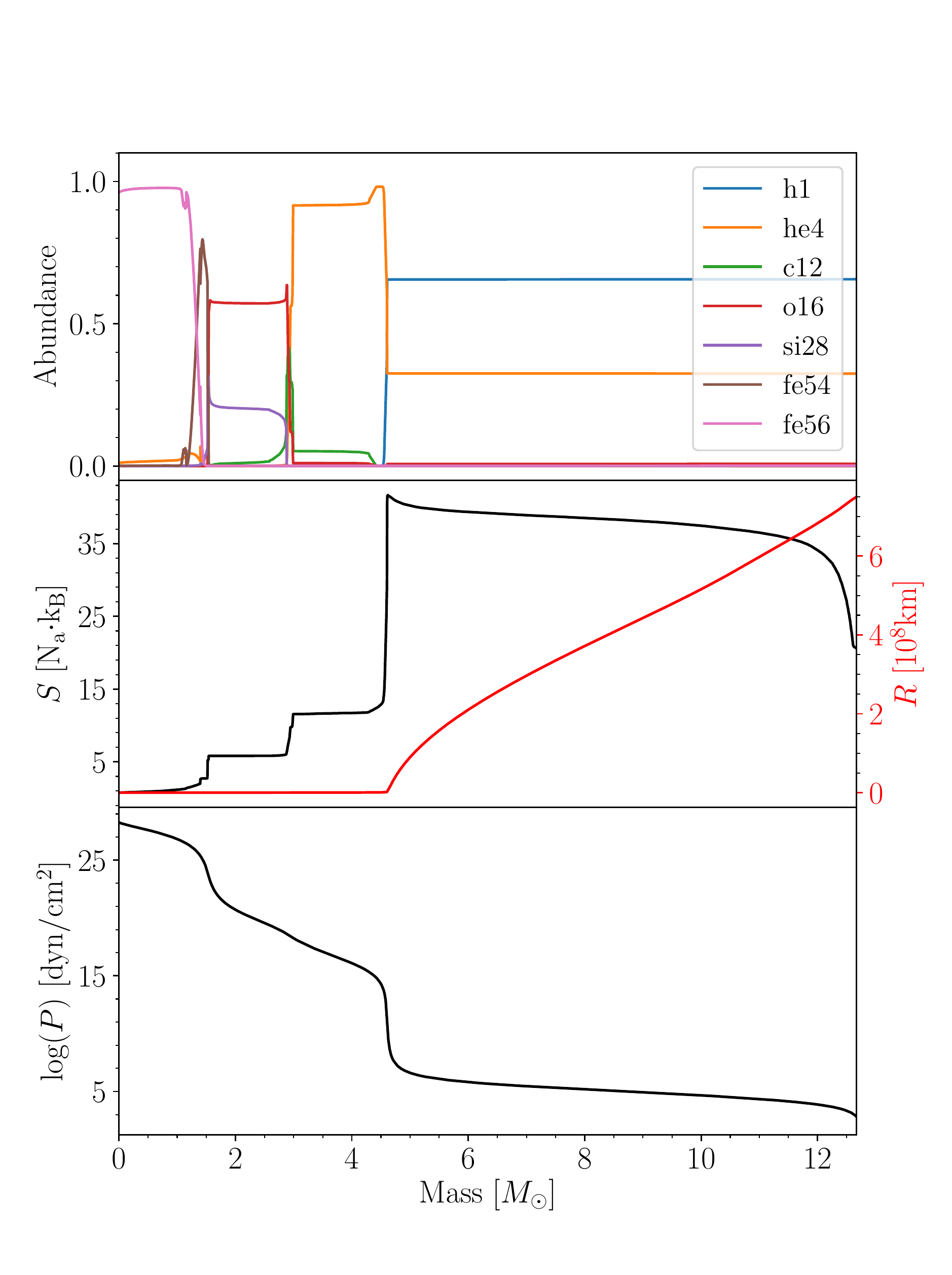}}
\caption{The profiles of the some quantities in the $M_{\rm ZAMS} = 15 M_\odot$ stellar model very close to core collapse. 
In the upper panel we present the composition of the main elements, in the middle panel we present the entropy (left axis, black line) and radius (right axis, red line), and in the lower panel we show the pressure, all as function of mass. }
\label{fig:Profiles15}
\end{figure}

\begin{figure} [ht] 
\hskip -0.6 cm
{\includegraphics[scale=0.4]{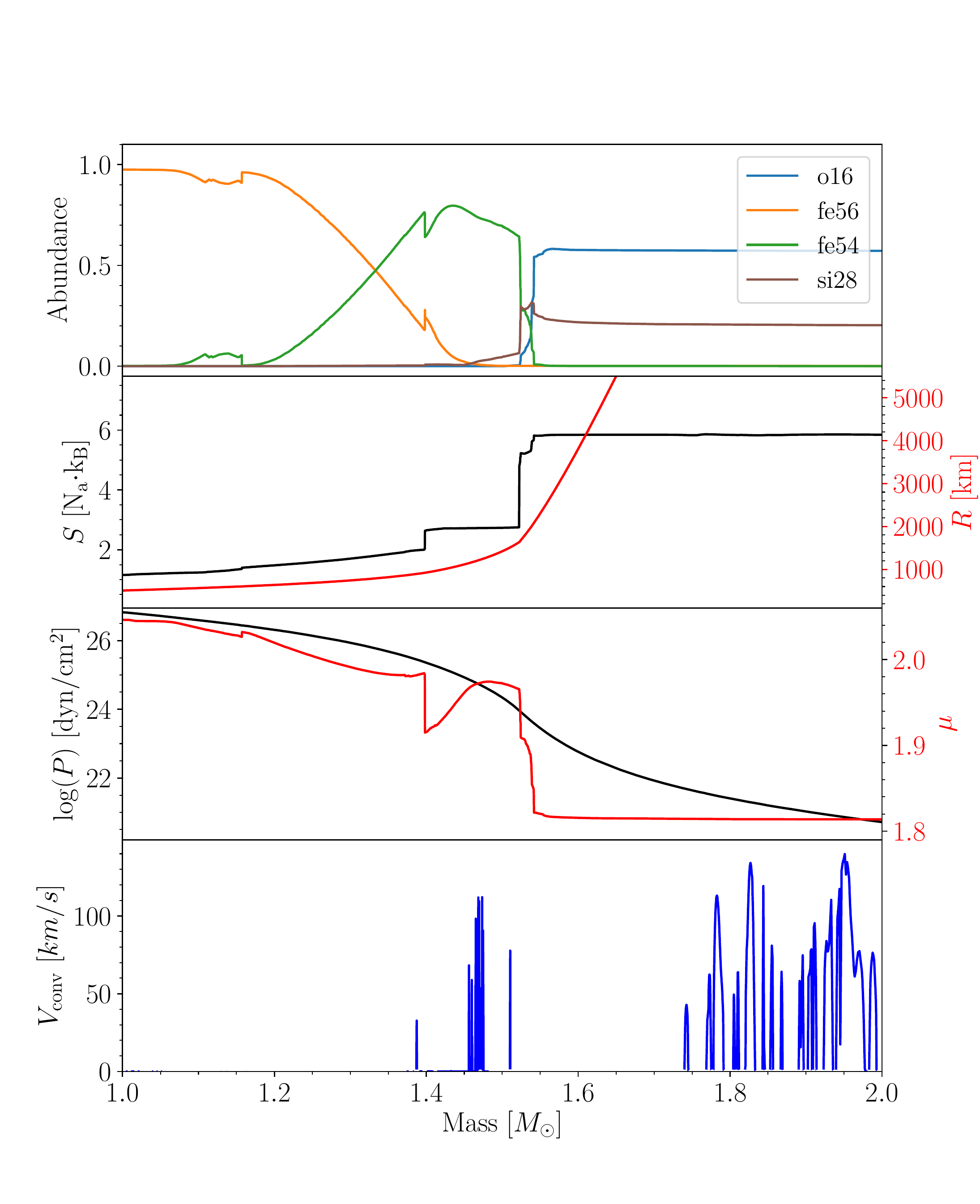}}
\caption{Like Fig. \ref{fig:Profiles15} but zooming in on the mass zone of $1M_\odot < m < 2 M_\odot$ and also adding the molecular weight in the third panel (right axis, red line) and a fourth panel of the convective velocity.
In the radiative zone the convective velocity is zero. }
\label{fig:Profiles15zoom}
\end{figure}

\begin{figure} [ht] 
\hskip -0.20 cm
{\includegraphics[scale=0.42]{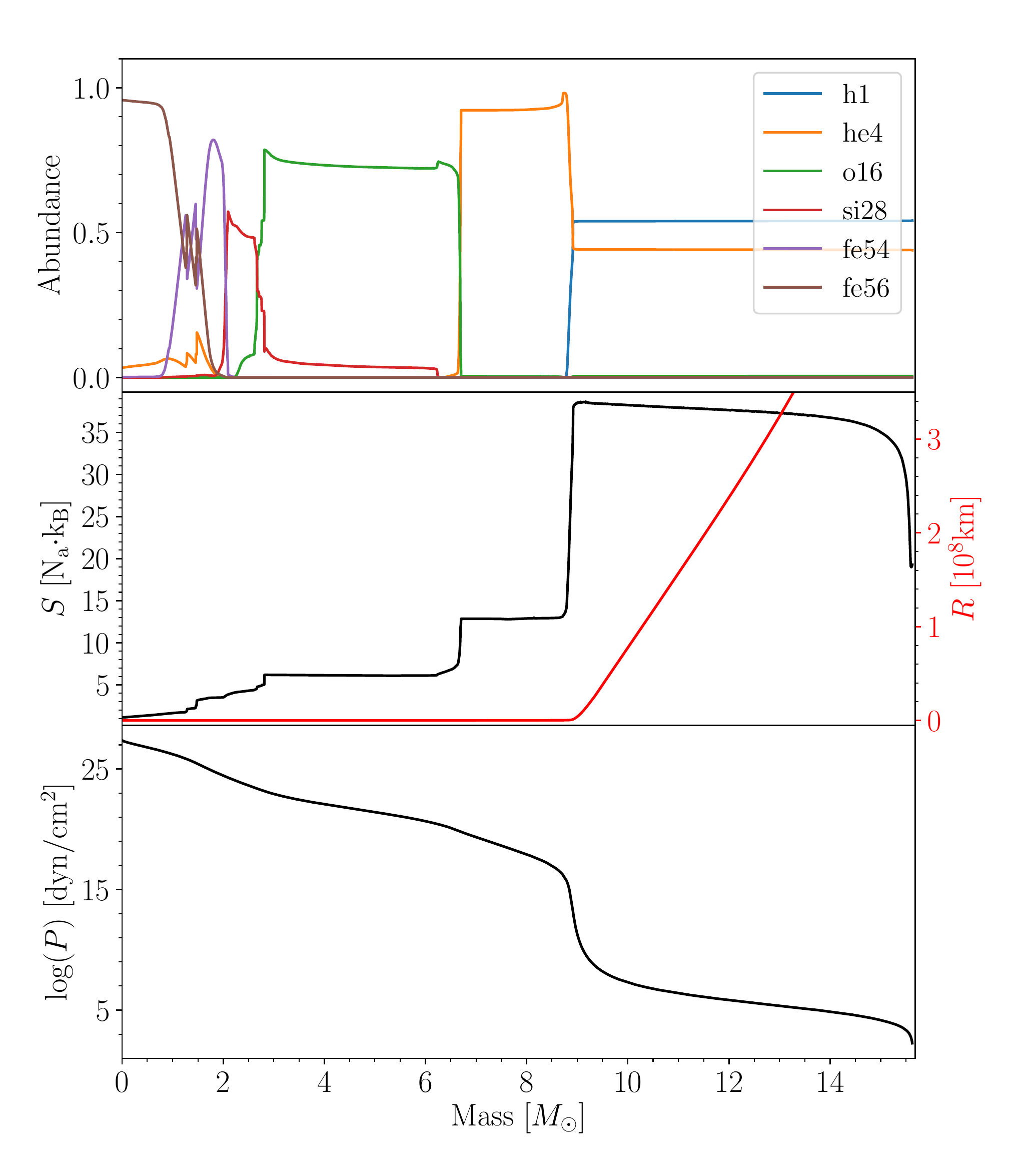}}
\caption{Like Fig. \ref{fig:Profiles15} but for the $M_{\rm ZAMS} = 25 M_\odot$ stellar model.
 }
\label{fig:Profiles25}
\end{figure}

\begin{figure} [ht]
\hskip -0.60 cm
{\includegraphics[scale=0.4]{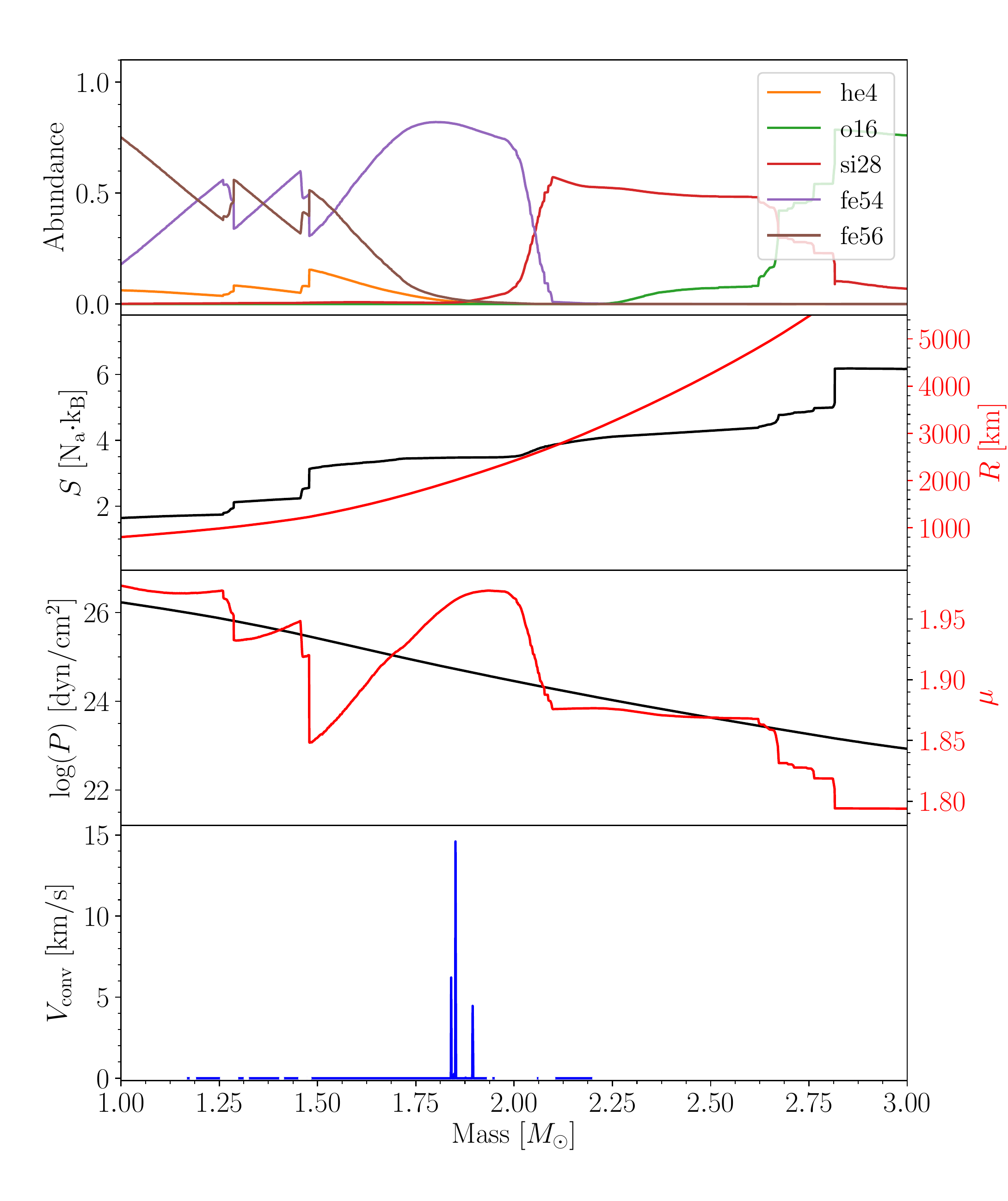}}
\caption{Like Fig. \ref{fig:Profiles15zoom} but for the $M_{\rm ZAMS} = 25 M_\odot$ stellar model and zooming in on the mass zone of $1 < m < 3 M_\odot$.
}
\label{fig:Profiles25zoom}
\end{figure}

In Fig. \ref{fig:Profiles15} we present the profiles of some relevant quantities in the model with an initial mass of $M_{\rm ZAMS} = 15 M_\odot$.
In Fig. \ref{fig:Profiles15zoom} we zoom in on the mass zone of $1 < m < 2 M_\odot$. 
In Figs. \ref{fig:Profiles25} and \ref{fig:Profiles25zoom} we present the same profiles but for a stellar model with an initial mass of $M_{\rm ZAMS} = 25 M_\odot$ and zooming in on the mass zone of  $1 < m < 3 M_\odot$.
Relevant to us is the convective zone at the outskirts of the iron core, and the steep entropy rise in the radiative zones above the convective zone. Namely, the mass zone of $1.4 M_\odot \la m \la 1.55 M_\odot$ in the $M_{\rm ZAMS} = 15 M_\odot$ model and the mass zone of $1.85 M_\odot \la m \la 2.8 M_\odot$ in the $M_{\rm ZAMS} = 25 M_\odot$ model. The structure is complicated due to several thin convective shells and an entropy rise in the mass zone just inner to $m=1.4 M_\odot$ ($m=1.85 M_\odot$) in the $M_{\rm ZAMS} = 15 M_\odot$ ($M_{\rm ZAMS} = 25 M_\odot$) model. Since at early times there were convective zones inner to that region, it is possible that this region also stores some magnetic fields from earlier dynamo activities.

\section{Storing magnetic fields}
\label{sec:results}

\begin{figure}  %
\hskip -0.5 cm
{\includegraphics[scale=0.41]{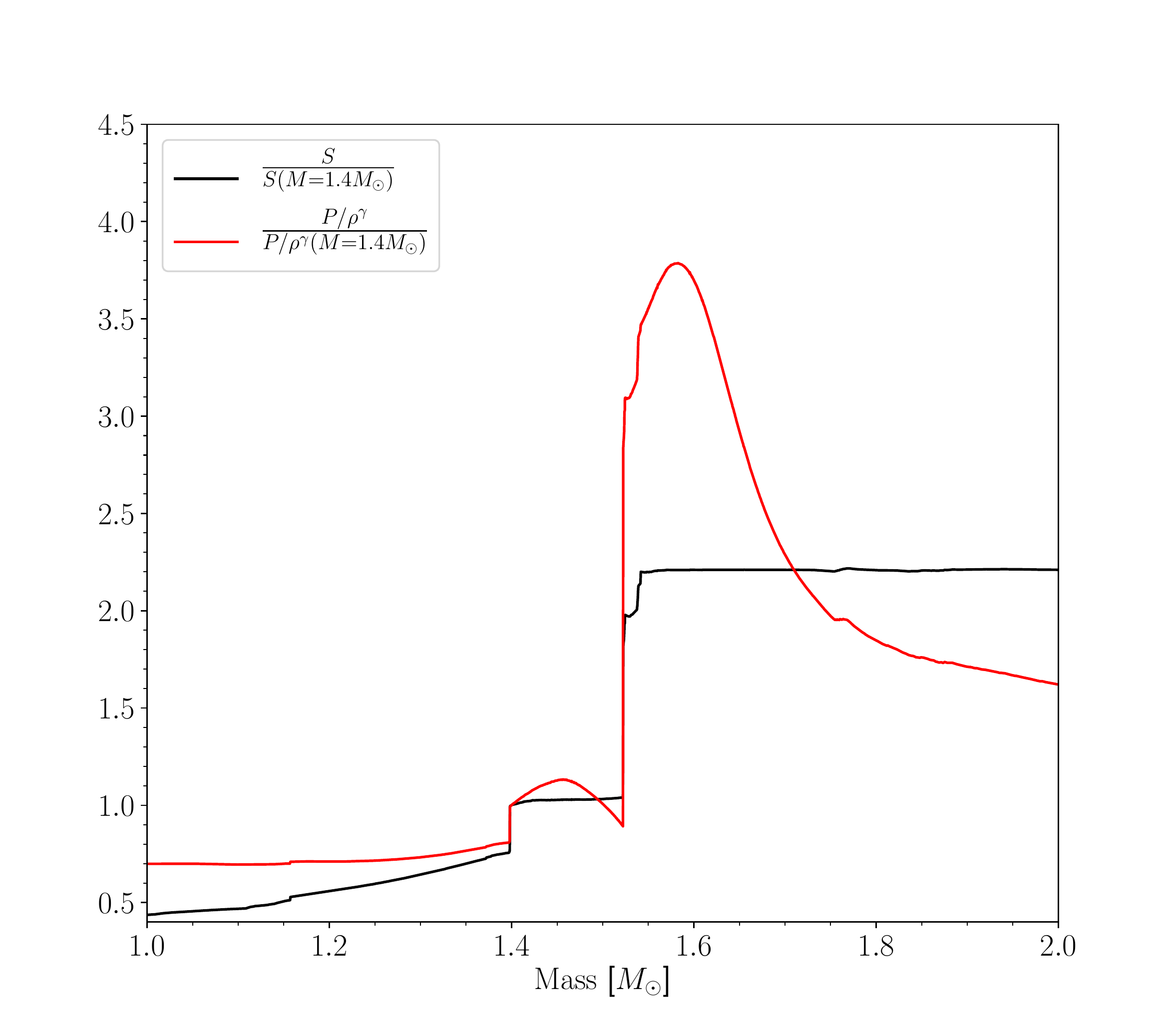}}
\caption{Scaled Entropy and scaled $S_{\gamma} \equiv P\rho^{{-\gamma}_{\rm ad}}$ profiles in the relevant region for NS formation for our $M_{\rm ZAMS}=15 M_\odot$ stellar model. We scale both quantities by their value at mass coordinate $m=1.4 M_\odot$. }
\label{fig:EntropyRatio15}
\end{figure}

In Fig. \ref{fig:EntropyRatio15} we present for the $M_{\rm ZAMS} = 15 M_\odot$ stellar model the scaled entropy and the scaled quantity $S_{\gamma} \equiv P\rho^{{-\gamma}_{\rm ad}}$ in the relevant region for NS formation, where $\gamma_{\rm ad}=\Gamma_1 \equiv (d \ln P / d \ln \rho)_{\rm s}$, i.e., the derivative is taken at a constant entropy. We scale both quantities by their value at the mass coordinate $m=1.4 M_\odot$. From Fig. \ref{fig:Profiles15zoom} we learn that the convective region where the dynamo might act efficiently is around $m \simeq 1.45 - 1.5 M_\odot$.
From that region the entropy increases by a factor of about 2 at $m \simeq 1.6 M_\odot$.

We can now substitute this ratio into equation (\ref{eq:equilibrium3}) and for the definition of $\beta_0 \equiv ( {P_{\rm thermal}}/{P_B} )_{t0}$ (equation \ref{eq:beta}), and obtain the maximum magnetic field that the dynamo can amplify and that the radiative zone can store 
\begin{equation}
{B}_{{\rm max}, 0}  \simeq (8 \pi )^{1/2} 
\left[
\left(  \frac{S_\gamma(r_{\rm max})}{S_{\gamma 0}} \right)^{1/1.3} -1 
\right]^{1/2}  P^{1/2}_{{\rm thermal}, 0},
\label{eq:Bfinal}
\end{equation}
where the pressure is taken at the dynamo location, i.e., at the relevant convective zone. For our $15 M_\odot$ model the relevant pressure is $P_{{\rm thermal},0} \simeq 4 \times 10^{24} \erg \cm^{-3}$, the value inside the square parenthesis in equation (\ref{eq:Bfinal}) is about $0.7$, and we find that  ${B}_{{\rm max}, 0}  \simeq 8 \times 10^{12} \G$. This magnetic field is about the value of the local magnetic field that \cite{Wheeleretal2015} found in their calculations of a dynamo above the iron core. Our new finding is that such magnetic fields can exist not only locally in the dynamo zone itself, but they can also be stored in some radiative and more extended regions above the dynamo zone. 

\begin{figure}  %
\vskip -0.0 cm
\begin{tabular}{cc}
\hskip -0.8 cm
{\includegraphics[scale=0.5]{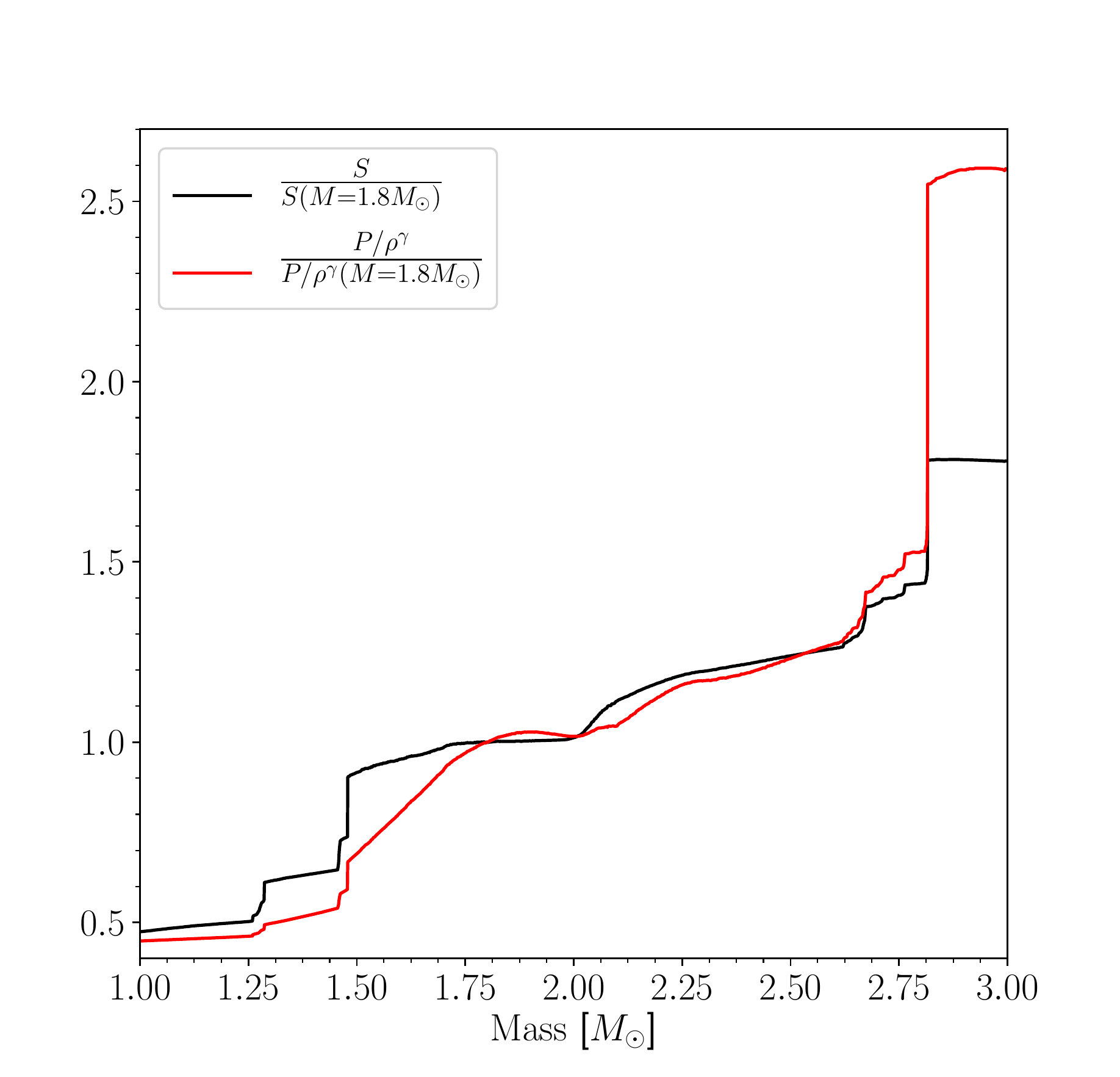}}
\end{tabular}
\caption{Like Fig. \ref{fig:EntropyRatio15} but for our $M_{\rm ZAMS}=25 M_\odot$ stellar model, and quantities scaled by their values at $m=1.8M_\odot$. }
\label{fig:EntropyRatio25}
\end{figure}

In Fig. \ref{fig:EntropyRatio25} we present the same scaled quantities before collapse for our $M_{\rm ZAMS}=25 M_\odot$ stellar model. We scale both quantities by their value at the mass coordinate $m=1.8 M_\odot$.  Repeating the calculation above but for the $M_{\rm ZAMS}=25 M_\odot$ model just before core collapse and for the radiative zone above $m\simeq 1.85 M_\odot$, we find ${B}_{{\rm max}, 0}  \simeq  10^{13} \G$, just slightly larger than the value for the $M_{\rm ZAMS}=15 M_\odot$ model.  
\section{Examining some assumptions}
\label{sec:assumptions}

We return to discuss our assumption of a constant molecular weight $\mu$. Earlier studies notice the role of a composition gradient (e.g., \citealt{MacDonaldMullan2004}, and \citealt{BraithwaiteSpruit2017, BrunBrowning2017} for recent reviews). \cite{MacDonaldMullan2004} argue that composition gradients prevent magnetic flux tubes that are generated in the core of massive stars to reach the surface of the star (for magnetic fields that are not above equipartition values).
Here we note the following. Firstly, the value of $\mu$ does not appear explicitly in equation (\ref{eq:equilibrium1}). In equation (\ref{eq:equilibrium3}) $\mu$ refers to its value at the origin of the flux tube. The value of $\mu$ does appear implicitly in equation (\ref{eq:equilibrium1}) through the ratio $P_e(r)/[\rho_e(r)]^{\gamma_{\rm ad}}$. We take the values of $P_e(r)$ and $\rho_e(r)$ directly from the stellar model, and so we implicitly take the variation of $\mu$ in the star into account.  

Secondly, in the relevant region of the core the molecular weight changes, but not by much. Its value decreases outward from $\mu = 1.98 $ at around $m=1.4 M_\odot$ to $\mu = 1.82$ at $m=1.6 M_\odot$ for the model with $M_{\rm ZAMS}=15 M_\odot$. 
For the model of $M_{\rm ZAMS}=25 M_\odot$ the value decreases from
$\mu = 1.98 $ at around $m=1.8 M_\odot$ to $\mu = 1.8$ at $m=3.0 M_\odot$.
  
For our assumption of $\delta= 0.5 \gamma_{\rm ad} \simeq 2/3$, the magnetic pressure and thermal pressure in the flux tubes change by the same ratio as the magnetic field rises. Let us assume that the value of $\mu$ of the envelope decreases by $\simeq 10 \%$. For the same pressure and temperature, the density would be lower by $\simeq 10 \%$. For the density of the flux tube to be equal to the new envelope density the magnetic pressure should be larger by $\simeq 0.1 P_e(r)$. As we discussed in section \ref{sec:results} we found here $P_{B0} \simeq 0.7 P_{\rm thermal,0}$ (equation \ref{eq:Bfinal}). The decreases in the molecular weight in the exercise here would require the magnetic field to be $P_{B0,\mu} \simeq 0.8 P_{\rm thermal,0}$. This is below the equipartition value. But as we said, we take the value of envelope pressure and density from the stellar model, which already includes the variation of the molecular weight. 

Another effect we did not study is the time variation just before collapse. We present the convective zones at a particular time shortly before collapse. We do note that the locations of the convective layers change rapidly before collapse (e.g., \citealt{Collinsetal2018}). A future study will have to explore the time evolution of the possible storing zones of magnetic fields. We here only note that even these rapid changes occur on time scales that are $>1000 \s$ (e.g., \citealt{Collinsetal2018}), which is more than three orders of magnitude longer than the dynamical time of the core. We conclude that there is sufficient time for a dynamo process to develop and build strong magnetic fields in the radiative zones.  

There are more effects that future studies will have to explore, including the three-dimensional structure of the convective zones in the core (e.g., \citealt{Couchetal2015, GilkisSoker2016}) and their boundaries with the radiative zones. 
\cite{Couchetal2015}, for example, find that the non spherical motion in the silicon convective zones are substantial. The influence of overshooting on dynamo operation should also be a subject of future studies. Our guess is that all these effects that add to the `non-quiet' core actually make the magnetic field amplification more likely. 

Finally, in the presence of differential rotation, the radiative zones themselves might sustain a magnetic dynamo \citep{Spruit2002}. This can further amplify the magnetic fields. In general, all the processes that add to the `noise' in the core in addition to the convection itself, are likely to further amplify magnetic fields (as second order effects), including rotational shear in radiative zones and their boundary with convective zones (e.g., \citealt{Zilbermanetal2018}), convective overshoot, some instabilities and the 3D vigorous convection motions.

\section{SUMMARY}
\label{sec:summary}

We study the radiative zone above the convective shells at the boundary of the iron core of massive stars just before they collapse. The steep entropy gradient in the radiative zone can store magnetic flux loops with relatively strong magnetic fields (section \ref{sec:storing}). We presented the steep entropy gradients in the relevant radiative zone in Fig. \ref{fig:Profiles15zoom} for a $M_{\rm ZAMS} =15 M_\odot$ stellar model and in Fig. \ref{fig:Profiles25zoom} for a $M_{\rm ZAMS} =25 M_\odot$ stellar model. 
In Figs. \ref{fig:EntropyRatio15} and  \ref{fig:EntropyRatio25} we present the relevant profiles scaled by their values at $m=1.4 M_\odot$ and $m=1.8 M_\odot$, respectively.
We then used the increase in the value of $S_\gamma$ to calculate the value of the maximum magnetic field that the radiative zone can store in flux tubes by equation (\ref{eq:Bfinal}).
The value of the magnetic field is that of the flux tube when it is formed at the convective shells at $m \simeq 1.5 M_\odot$ for the $M_{\rm ZAMS} =15 M_\odot$ stellar model and at $m \simeq 1.85 M_\odot$ for the $M_{\rm ZAMS} =25 M_\odot$ model. 

We found that magnetic flux tubes that emerge from the convective shells with magnetic fields of $B_0 \la B_{\rm max,0} \approx 10^{13} \G$ stop their buoyancy within the radiative zone. 

Our main result is therefore that the regions above the iron core not only produce locally strong magnetic fields in the convective shells (e.g.,  \citealt{Wheeleretal2015}), but that the radiative zone can store such magnetic fields in a more extended zone. 
 
When these magnetised zones flow toward the center as the core collapses, the flux tubes can serve as seeds for further magnetic field amplification by instabilities behind the stalled shock. The strong magnetic fields have a crucial role in launching jets in the jittering jets explosion mechanism \citep{Soker2018, Soker2019}.
During core collapse mass shells from radii of $r_i \approx {\rm few} \times 1000 \km$ fall radially inward and pass through the stalled shock at $r_s \simeq 100 \km$. This inflow amplifies only the radial component of the magnetic field. For an initial random magnetic field the inflow increases the average magnetic field intensity by $\approx (1/3)^{1/2} (r_i/r_s)^2$, which can amount to a factor $>100$.

We conclude that when the pre-collapse steep entropy rise zone above the iron core crosses the stalled shock at $r_s \simeq 100 \km$ it can bring with it magnetic fields of up to $B_{\rm sh} \approx 100 B_{\rm max,0} \approx {\rm few} \times 10^{14} \G$. 
In many cases the real stored magnetic field will be weaker, depending on the operation of the dynamo. We note that \cite{Endeveetal2012} start their spiral standing accretion shock instability simulations with an average magnetic field intensity at $r=100 \km$ of only $10^9-10^{12} \G$. We allow for much stronger magnetic fields, and hence further support the claim for the necessity to include magnetic fields in simulating the explosion of CCSNe \citep{Soker2018}.

We thank Avishai Gilkis for his help in proper adjustment of the stellar parameters in MESA for the evolution of massive stars.
We thank an anonymous referee for useful comments. 
This research was supported by the Israel Science Foundation.

\label{lastpage}
\end{document}